\documentclass[journal,twocolumn]{IEEEtran}
\usepackage{epsfig,makeidx,color,subfigure}
\usepackage{amsmath,amssymb,bbm}
\usepackage{cite,graphicx,lipsum}
\usepackage{enumerate}
\usepackage{empheq}
\usepackage[switch,pagewise]{lineno}
\usepackage{hyperref}
\hypersetup{
        colorlinks = true,
        citecolor=blue,
}
\def\cM{{\cal M}}

\def\uE{{\mathbb E}}

\newtheorem{mylemma}{\bf Lemma} % [section]
\newtheorem{myproposition}{\it Proposition} % [section]

\def\be{ \begin{equation} }
\def\ee{ \end{equation} }
\def\bea{ \begin{eqnarray} }
\def\eea{ \end{eqnarray} }

\def\b0{{\bf 0}}

\ifCLASSOPTIONonecolumn
  \newcommand{\figwidth}{0.40\columnwidth}
\else
  \newcommand{\figwidth}{0.80\columnwidth}
\fi
\usepackage{graphicx}
\usepackage{booktabs}
\usepackage{graphicx}
\usepackage[normalem]{ulem}

\begin{document}

\title{Random Access-based Multiuser Computation Offloading 
for Devices in IoT Applications}

\author{Jinho Choi\\
\thanks{The author is with
the School of Information Technology,
Deakin University, Geelong, VIC 3220, Australia
(e-mail: jinho.choi@deakin.edu.au).
This research was supported
by the Australian Government through the Australian Research
Council's Discovery Projects funding scheme (DP200100391).}}
\maketitle

\begin{abstract}
In various Internet-of-Things (IoT) applications, a number of devices and sensors are used to collect data sets. As devices become more capable and smarter, they can not only collect data sets, but also process them locally. However, since most devices would be limited in terms of computing power and energy, they can take advantage of offloading so that their tasks can be carried out at mobile edge computing (MEC) servers. 
In this paper, we discuss computation offloading  for devices in IoT applications. In particular, we consider users or devices with \emph{sporadic} tasks, where optimizing resource allocation between offloading devices and coordinating for multiuser offloading becomes inefficient. Thus, we propose a two-stage offloading approach that is friendly to devices with sporadic tasks as it employs multichannel random access for offloading requests with low signaling overhead. 
The stability of the two-stage offloading approach is considered with methods to stabilize the system. We also analyze the latency outage probability as a performance index from a device perspective. 
\end{abstract}


\ifCLASSOPTIONonecolumn
\baselineskip 23pt
\fi

\section{Introduction}

It is expected to have more computing intensive tasks at user equipment (UE) such as smart phones, while UEs are limited in terms of computing power and energy source. Thus, as in \cite{Yang08} \cite{Kumar10}, computation offloading can be considered when computation cost exceeds communication cost to save UE's energy and exploit the powerful computing power of cloud servers. In cellular systems, mobile edge computing (MEC) \cite{Mach17} \cite{Abbas18} can be employed so that  MEC servers integrated into base stations (BS) can provide computing power without excessive latency thanks the physical proximity.

%Although new mobile devices are more and more powerful in terms of central processing unit (CPU), even these may not be able to handle the applications requiring
%huge processing in a short time.

Multiple UEs or users can co-exist in a cell and want to perform computation offloading simultaneously. In this case, it is necessary to consider resource allocation for multiuser MEC offloading. 
In \cite{Barbarossa13}, an optimization problem is formulated to minimize the total transmit power of multiple users subject to constraints on latency. Since mobile devices have limited energy sources, in \cite{You17}, using an objective function based on the total energy consumption, an optimization problem is formulated and solved. In multiuser MEC offloading, since multiple users are to compete for the shared resource (especially, limited radio bandwidth), in \cite{Chen16}, a non-cooperative game is formulated and a distributed algorithm is devised. Compared to the approaches in \cite{Barbarossa13} \cite{You17}, the approach in \cite{Chen16} is suitable for the cases where a BS 
cannot obtain the necessary information to solve optimization problems.  For example, some users may not want to share their computation capacity information (e.g., the number of central processing unit (CPU) cycles per second) with the BS or cannot predict available computation capacity due to other possible tasks in the future. 

Multiuser MEC offloading has been extended to incorporate other aspects. 
For example,
in \cite{Hu18} \cite{Feng19} \cite{Min19}, offloading is considered with wireless power transfer, and optimization problems are extended to include energy harvesting. In addition,  multiuser MEC offloading is  extended in \cite{Zhou21} to deal with different user requirements in heterogeneous networks, and in \cite{Anajemba20} for cooperative offloading in a heterogeneous network consisting of small-cell BSs and wireless relays.
%In \cite{Non-orthogonal multiple access (NOMA) \cite{Ding_CM} \cite{Choi17_ISWCS} based 

In the paper, we consider a different setup from a conventional one, e.g., the setup used in \cite{You17} \cite{Chen16}. In particular, we consider the case that users are devices that have computation tasks at random times in Internet-of-Things (IoT) applications. 
Smart sensors and devices may not only collect data, but also process data with limited computing power. Thus, there are devices that perform tasks on their own, while some devices may choose 
to offload sporadic tasks to a BS that is associated with an MEC server so that the MEC server can perform tasks. 
In most IoT applications, traffic is considered sporadic \cite{Bockelmann16} \cite{Metzger19}, and events of interest, especially in emergency IoT applications, are often rare and sporadic \cite{Sisinni20}.
Thus, as in \cite{Patel17}, in order 
to speed up an analytic task in certain IoT applications such as real-time video analytics, cloud gaming, and smart factories, 
offloading can considered for parallelization.
Of course, offloading decisions can depend on many factors and each device and task as in conventional offloading approaches. 
The main difference, however, is that IoT devices' tasks that require offloading occur sporadically and are relatively small, so 
optimizing resource allocation between offloading devices and coordinating for multiuser offloading by the BS centrally becomes inefficient (as in most conventional approaches), because the time and signaling overhead to gather the necessary information offsets the benefits of offloading.
This is quite similar to machine-type communication (MTC) \cite{Bockelmann18} where devices have sporadic traffic and become active to transmit data packets at random times in a number of IoT applications \cite{Ding_20Access}. 
Thus, unlike human-type communication (HTC), non-coordinated transmission schemes such as random access are employed for MTC due to low signaling overhead without any specific channel resource allocations for multiple devices with sporadic traffic. For example, two-step random access \cite{Kim21} \cite{Choi_MWC} is an MTC scheme introduced in 5th generation (5G) for a large number of devices with sporadic traffic.

Note that in \cite{Toma14}, computation offloading for sporadic tasks is studied for a fixed number of users without any limitation on accessing wireless channels. In IoT applications, however, due to a large number of devices with limited bandwidth, it is necessary to study offloading along with how efficiently a number of devices utilize the shared radio resources. In this paper, as mentioned earlier, the proposed approach leverages the two-step
random access in 5G to access the shared radio resource, making it suitable for IoT applications.

To support offloading for devices with sporadic tasks, we propose a two-stage approach. In the first stage, devices send requests through (multiple) random access channels to a BS. If the BS can successfully receive requests without collisions, it schedules devices' uploading the input data for offloading. In the second stage,  offloading enabled devices send their input data according to a given schedule. The proposed approach has the advantage that it can support with low signaling overhead when device offloading is needed sporadically. 
%In conventional approaches where radio resource allocation is to be optimized for multiuser MEC offloading (e.g., \cite{Barbarossa13} \cite{You17}), the signaling overhead to coordinate multiple users' offloading tasks may result in undesirable latency, and the amount of time for signaling can exceed the latency limit.
%As a result, they are not suitable when devices need offloading for sporadic tasks with certain latency constraints, which we consider in this paper.

The main contributions of the paper can be summarized as follows. 
\begin{itemize}
\item To support offloading by devices with sporadic tasks, a two-stage offloading approach with low signaling overhead is proposed, where each device makes a decision on offloading locally. This approach leverages  two-step random access, a new MTC protocol in 5G, and suitable for IoT devices.
Note that 
conventional offloading methods, in which the BS centrally optimizes radio resources, becomes inefficient due to  devices' sporadic tasks (because the signaling overhead to coordinate multiple users' offloading tasks may be excessive and result in undesirable latency).

\item The stability of the two-stage offloading approach is studied with control approaches for a stable system, which is important as the proposed approach is a distributed system.

\item The latency outage probability is analyzed to see the performance from a device perspective for statistical guarantees on task completion times.
\end{itemize}

The rest of the paper is organized as follows. In Section~\ref{S:SM}, we present a system model for the two-stage offloading approach with details for each stage. A system perspective for the two-stage offloading approach is discussed in Section~\ref{S:SP}, along with stability and methods to ensure a stable system by controlling key parameters. We also discuss a device perspective in Section~\ref{S:LO} with the analysis of latency outage probability. Simulation results are presented in Section~\ref{S:Sim}. We finally conclude the paper with some remarks in Section~\ref{S:Con}.

%\subsubsection*{Notation}
%Matrices and vectors are denoted by upper- and lower-case
%boldface letters, respectively.
%The superscript $\rT$
%denotes the transpose.
%For a set $\cA$, $|\cA|$ denotes the cardinality of $\cA$.
%$\uE[\cdot]$ and ${\rm Var}(\cdot)$
%denote the statistical expectation and variance, respectively.
%$\cC \cN(\ba, \bR)$ represents the distribution of circularly symmetric complex Gaussian (CSCG) random vectors with mean vector $\ba$ and covariance matrix $\bR$.

The definitions of key parameters are as follows.

\begin{tabular}{ll}
$K$: & number of active devices (per round) \cr
$W$: & number of contending devices \cr 
$S$: & number of offloading devices \cr 
$M$: & number of random access channels  \cr 
$B$: & total system bandwidth \cr 
$\Delta$: & interval of random access round \cr 
$U_{\rm max}$: & maximum size of input for offloading \cr 
\end{tabular}

\section{System Model}  \label{S:SM}

Suppose that a system consists of one BS and multiple devices. Each device may have a (computation) task to be performed and can choose computation offloading so that its task can be performed at an MEC server connected to the BS.

For computation offloading, we consider the following two-stage approach:
\begin{enumerate}
    \item Stage 1: Once a device decides to offload its computing task, it sends a request to the BS using multichannel random access. To this end, grant-free or two-step random access \cite{Kim21} \cite{Choi_MWC} is used, where the device sends a preamble and a short packet of offloading request. This packet includes metadata such as the size of the input data in bits to be uploaded for offloading.
    \item Stage 2: The BS sends the feedback signal of positive or negative acknowledgement (ACK or NACK) to inform the success of transmission of request with the time to start uploading the input data through the dedicated uplink channel in a scheduled time division multiple access (TDMA) manner.
\end{enumerate}

As mentioned earlier, devices have sporadic tasks, which makes resource
optimization for offloading with specific devices at any given moment difficult. Therefore, when a device needs offloading, it is appropriate to immediately request offloading by sending packets of offloading requests to the BS. Due to multiple devices that can send request simultaneously, in stage 1, we consider multichannel random access, especially two-step random access proposed in 5G. 
In two-step random access, the payload is fixed in size and short \cite{Choi_MWC}, making it suitable for sending offloading requests to devices with sporadic tasks. The BS can schedule uploads from the devices that successfully send requests without collisions.

For the two-stage approach, the total uplink system bandwidth, $B$, is divided into two groups as follows:
\be 
B = B_{\rm o} + B_{\rm a},
    \label{EQ:BBB}
\ee 
where $B_{\rm a}$ is the bandwidth allocated for random access to perform the first stage and $B_{\rm o}$ is the bandwidth allocated for uploading in the second stage. The channel of bandwidth $B_{\rm o}$ is referred to as the offloading channel (OC), while that of bandwidth $B_{\rm a}$ the random access channel (RAC).
It is further assumed that 
there are $M$ multiple sub-channels within the RAC so that $B_{\rm a} = M b$, where $b$ is the bandwidth of one channel for RAC and 
$$
M \in \cM = \left\{1, \ldots, \frac{B}{b} \right\}, \ B \gg b.
$$
Note that if $M = 1$, it is single-channel ALOHA \cite{BertsekasBook}. 
It is assumed that a random access round is executed periodically every $\Delta$ seconds.

\begin{figure}[thb]
\begin{center}
\includegraphics[width=\figwidth]{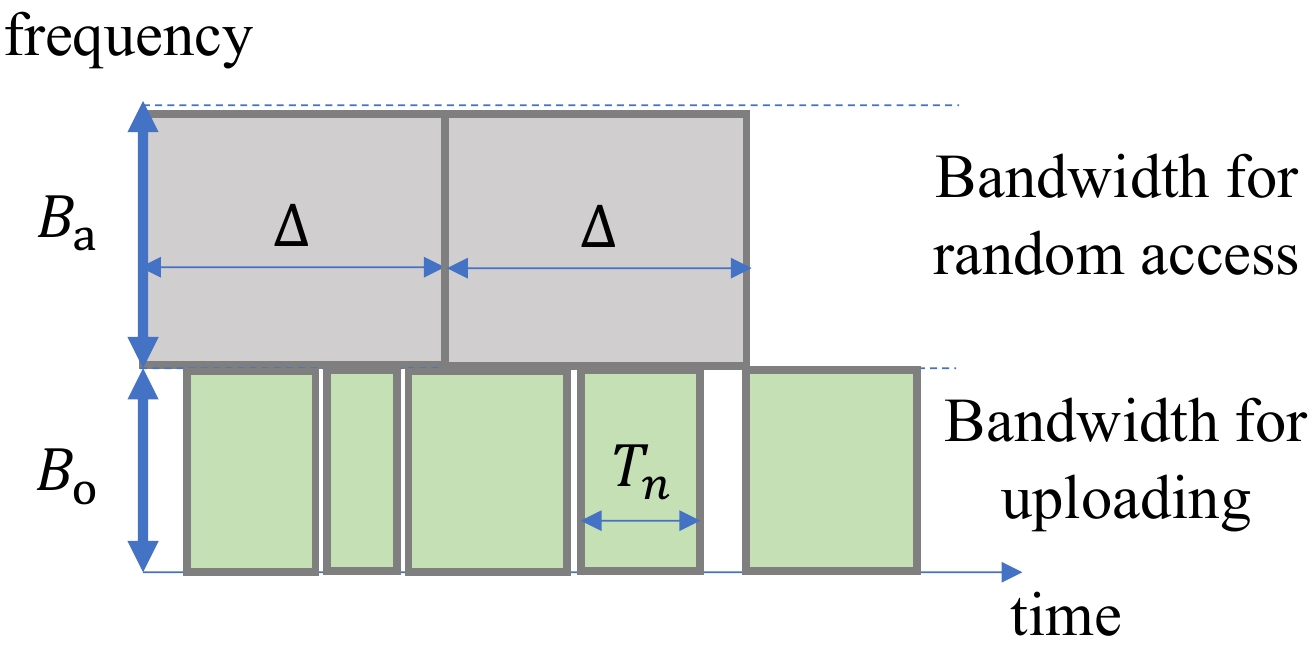}
\end{center}
\caption{The bandwidth allocation for two stages that can take place simultaneously. The scheduled TDMA for Stage 2 with different size of input data, denoted by $T_n$, will be explained below.}
        \label{Fig:Sys}
\end{figure}

In Fig.~\ref{Fig:Sys}, we illustrate the bandwidth allocation for two stages. An active device to offload its computing task performs Stage 1 in the RAC (i.e., two-step random access to send an offloading request packet). If this request is accepted by the BS, the BS sends ACK and schedules the uploading so that the device can upload its input data in scheduled TDMA manner in the OC of bandwidth $B_{\rm o}$ as Stage 2. 
By keeping the OC and RAC separate, Stages 1 and 2 can be performed concurrently for devices with sporadic tasks occurring at different times.
From this, devices do not need to synchronize, nor do they need to have the same size of input data for computation offloading.

\begin{figure}[thb]
\begin{center}
\includegraphics[width=\figwidth]{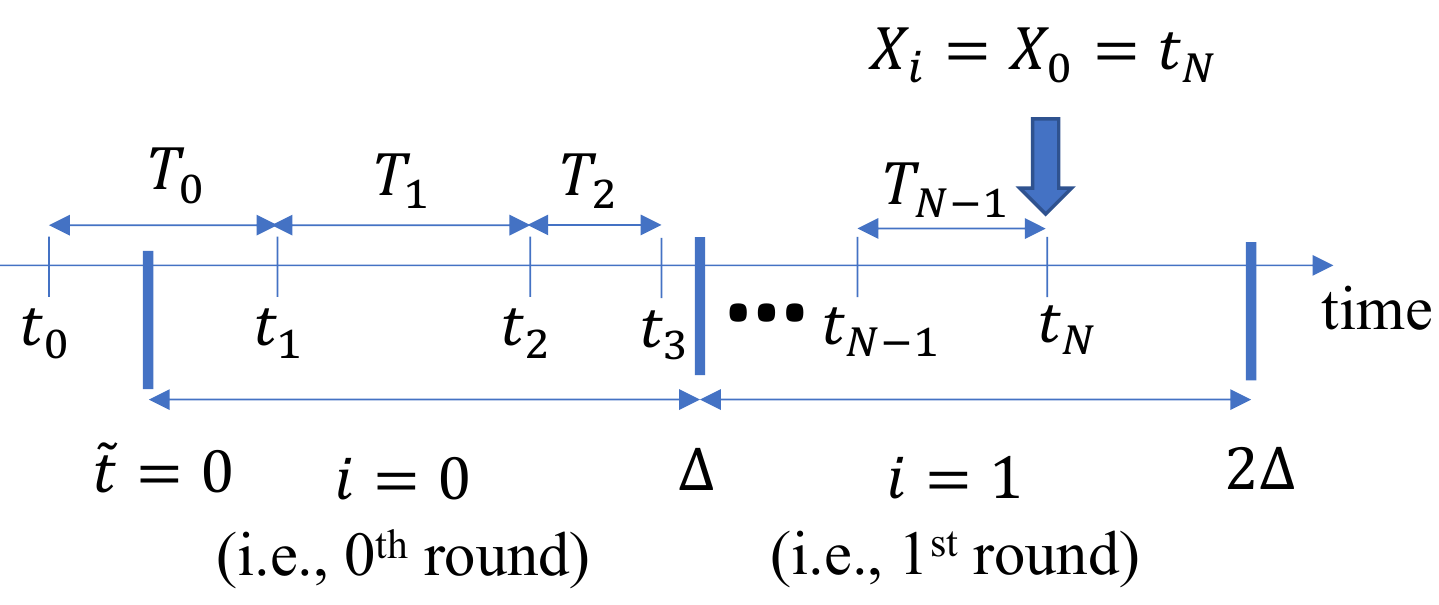}
\end{center}
\caption{Random access rounds and existing tasks at time $\tilde t = 0$
(i.e., the 0th random access round), where $t_N = X_0$.}
        \label{Fig:times}
\end{figure}

The time for random access rounds is denoted by $\tilde t \in \{i \Delta\}$, where $i$ is the integer time index for random access rounds, as shown in Fig.~\ref{Fig:times}. We assume that the current time for random access round is $\tilde t = 0$ or $i =0$, and  there are $N$ existing offloading tasks, where $N \ge 0$. 
Denote by $t_n$ the time when the uploading for scheduled offloading task $n$ begins in a scheduled TDMA manner. Since $t_0$ is the time that the earliest existing task at time $\tilde t = 0$, we assume that $t_0 < 0 < t_1$ as shown in Fig.~\ref{Fig:times}. 
Then, we have
\be 
t_{n+1} = t_n + T_n,
\ee 
where
\be 
T_n = \frac{U_n}{B_{\rm o} \log_2 (1 + \gamma_n) } .
    \label{EQ:T_n}
\ee 
Here, $U_n$ is the number of bits to be uploaded for offloading 
and $\gamma_n$ the SNR of the device associated with existing offloading task $n$.

We assume that the BS broadcasts the time that all the existing uploads are complete in each random access round, i.e., $t_N$ is available as the state of system at the devices in the random access round at $\tilde t = 0$
according to Fig.~\ref{Fig:times}. 
For convenience, %at the $i$th random access round, 
denote by $X_i$ 
the time that all the existing uploads scheduled prior to the $i$th random access round finish. For example, as shown in Fig.~\ref{Fig:times},
when $i = 0$, 
$t_N$ becomes $X_i = X_0$. Throughout the paper, $X_i$ is referred to as the state variable.

\section{System Perspective: Stability and Offloading Criteria} \label{S:SP}

%In this section, we discuss key  conditions to ensure the system in Section~\ref{S:SM} stable under certain assumptions.

In this section, we discuss the random access based offloading approach
in Section~\ref{S:SM} from a system perspective. 
Under certain assumptions, the stability of the system will be considered to avoid the upload time from continuously growing.

\subsection{Stability and Main Assumptions}

For a given round of random access of duration $\Delta$, say round $i$,
denote by $D_i$ the total 
upload time during which the devices that have allowed offloading through stage 1 (random access round) can upload data through OC.
Thus, it can be shown that 
$$
D_i = X_{i+1} - X_i,
$$
which is known to the devices once $X_{i+1}$ is sent by the BS (clearly, $D_i$ is unavailable to devices at the beginning of random access round $i$).
While the duration of random access round is fixed, the total upload time can vary as it depends on the number of the offloading devices\footnote{A device that successfully sends its offloading request in stage 1 and its upload for offloading is scheduled in stage 2 is called an offloading device.}, the sizes of the data to be uploaded, and channel conditions. 

Suppose that there are $Q$ rounds. It can be seen that the system may not suffer from a significant uploading delay if 
\be 
\sum_{i=0}^{Q-1} D_i \le Q \Delta.
\ee 
If $D_i$ is independent and identically distributed (iid), due to the law of large numbers \cite{Mitz05}, we have 
\be 
\lim_{Q \to \infty} \frac{1}{Q}\sum_{i=0}^{Q-1} D_i = \uE[D_i] \le \Delta,
    \label{EQ:stable}
\ee 
where $\uE[\cdot]$ represents the statistical expectation.
Consequently, \eqref{EQ:stable} can be seen as a (asymptotic) stability condition for a large $Q$.

To find $\uE[D_i]$, we consider one random access round and omit the index of round, $i$, in the rest of this section.
For a given round,
suppose that there are $K$ active devices with computation tasks at time $\tilde t = 0$.
Denote by $U_{(k)}$ and $\gamma_{(k)}$ the number of bits to be uploaded and the SNR of device $k$, respectively.
Throughout this section, we will consider the following assumptions.

\begin{itemize}
\item[{\bf A1})] The number of new active devices with computation tasks in each round of length $\Delta$, i.e., $K$, follows a Poisson distribution with parameter $\uE[K] = \lambda$. For a finite number of devices in a cell, this assumption is an approximation. In particular, if the total number of devices is $G$ and each device can have a task sporadically with a probability of $\epsilon$, $K$ will follow a binomial distribution, i.e., 
$K \sim \Pr(K = k) = \binom{G}{k} \epsilon^k (1-\epsilon)^{G-k}$. When $G\gg 1$ and $\epsilon$ is sufficiently small, $K$ can be well approximated by a Poisson random variable with mean $\lambda = G \epsilon$ \cite{Mitz05}.

%Since there are also active devices in the previous rounds that decide to re-transmit, $K \ge N$ (recall that $K$ is the number of active devices at time $\tilde t = 0$).
%\item[{\bf A2})]    For given $S$, the uploading order is uniformly decided. That is, $\Pr(S(k) = s) = \frac{1}{S}$, $s = \{1,\ldots, S\}$.

\item[{\bf A2})] The $U_{(k)}$'s are independent and  follow the same distribution that is an exponential distribution with $\uE[U_{(k)}] = \frac{1}{\mu}$, i.e.,
\be 
U_{(k)} \sim {\rm Exp}(\mu) = f_U (u) = \mu e^{-\mu u}, \ u \ge 0.
\ee 
Here, $\frac{1}{\mu}$ represents the average size of input. 
\item[{\bf A3})] The SNRs, $\gamma_{(k)}$, are independent 
and $\uE\left[ \frac{1}{\gamma_{(k)}} \right]$ is finite. For example,
$\gamma_{(k)}$ follows a truncated exponential distribution, i.e.,
\be 
f(\gamma_{(k)} = \gamma) = 
\left\{ 
\begin{array}{ll}
\frac{ \exp\left( - \frac{\gamma- \Gamma_{(k)}}{\bar \gamma } \right) }{\bar \gamma } , & \mbox{if $\gamma \ge \Gamma_{(k)}$;} \cr
0, & \mbox{o.w.} \cr 
\end{array}
\right.
    \label{EQ:gdist}
\ee 
Here, $\bar \gamma  = \uE[\gamma_{(k)}]$ and
$\Gamma_{(k)} > 0$ is the SNR threshold. 
%That is, a device with SNR below this threshold does not choose offloading as its uploading time might be long.
%As a result, $T_{(m)}$, $m= 1,\ldots, S(k)-1$, in \eqref{EQ:Yk} are iid.
\end{itemize}

To model sporadic tasks generated at devices, the assumption of {\bf A1} is considered. In general, it is expected that $\lambda$ increases with $\Delta$. The size of input data varies from a task to another. Thus, in the assumption of {\bf A2}, an exponential distribution is considered. It is noteworthy that  the two assumptions of {\bf A1} and {\bf A2} result in an M/M/1 queue in queueing theory \cite{Kleinrock79}. 
The assumption of {\bf A3} is essential to avoid an excessive upload time due to a low SNR.  
In fact, if the SNR is low in stage 1, the device cannot send an offloading request packet, so it can be assumed that the SNR in stage 2 is high enough.

\subsection{Decision Criteria} 

For each active device to choose offloading, there should be a criterion. We can consider two different types of  criteria as follows.
\begin{itemize}
\item Energy-based criteria: For a given task, each active device can find the energy for local computing and that for offloading. If the energy for offloading is lower than that for local computing, it can choose to offload the task. In multiuser cases, optimization problems can be formulated and solved for radio resource allocation (e.g., \cite{Chen16} \cite{You17} \cite{Wang17}). To this end, devices need to send information to the BS including their computing power (i.e., CPU cycles per second), levels of energy for computing, and so on.

\item Latency-based criteria: Each active device needs to find the time to complete a task under  two possible scenarios: local computing and offloading. In multiuser cases, the device cannot exactly determine the time to complete task under offloading as there can be other devices. Thus, the devices need to send their information to the BS so that the BS can optimize and decide whether or not devices can offload. 
\end{itemize}

In this paper, we consider local decision for offloading at devices. In particular, in this section, a local decision rule based on the size of input data is considered, where active device $k$  chooses local computing 
if the size of input data, $U_{(k)}$, is greater than a threshold, denoted by $U_{\rm max}$, as the transmission time and the energy consumption to upload the input data can be too long and high, respectively.

\subsection{Mean of Total Uploading Time}

Let $W \ (\le K)$ be the number of devices that decide to send offloading requests among $K$ active devices, which are called the contending devices, with probability 
$q_{\rm o}$. This probability is referred to as the offloading probability. For given $U_{\rm max}$, according to the assumption of {\bf A2}, the offloading probability is given by
\begin{align}
q_{\rm o} &= \Pr(U_{(k)} \le U_{\rm max} ) \cr
& = 1 - e^{-\mu U_{\rm max}}. 
    \label{EQ:qo}
\end{align}
Then, for given $K$, the number of the contending devices has the following distribution:
\be 
\Pr(W=w\,|\,K) = \binom{K}{w} q_{\rm o}^w (1- q_{\rm o})^{K-w}.
\ee 
From this, according to the assumption of {\bf A1}, it can be shown that
\be 
W \sim {\rm Pois}(\lambda q_{\rm o}).
    \label{EQ:WP}
\ee 
Then, there are $W$ contending devices in stage 1. 
Let $S \ (\le M)$ be the number of contending devices that can successfully transmit their requests without collisions in stage 1, which are called the offloading devices. That is, $S$ stands for the number of the offloading devices.
Clearly, $S \le \min\{W,M\}$.
Finally, the total upload time is given by
\be 
D = \sum_{m=1}^S \tilde T_{(m)}, 
    \label{EQ:DX}
\ee 
where $\tilde T_{(m)}$ represents the upload time of the $m$th offloading device through OC, which is
\be 
\tilde T_{(m)} = \frac{\tilde U_{(m)}}{B_{\rm o} \log_2 (1 + \tilde \gamma_{(m)}) } .
    \label{EQ:tXm} 
\ee
Here, $\tilde U_{(m)}$ and 
$\tilde \gamma_{(m)}$ are the number of bits for offloading and the SNR of the $m$th offloading device, respectively.
Clearly, $\{\tilde T_{(m)}, \ldots, \tilde T_{(M)}\}$ is a subset of
$\{T_{(1)}, \ldots, T_{(K)} \}$.

\begin{mylemma} \label{L:1}
Under the assumptions of {\bf A1} -- {\bf A3} with an identical distribution of $\gamma_{(k)}$, the mean of the total upload time is given by
\begin{align} 
\uE[D] 
& = \uE[S] \uE[\tilde T_{(m)}] \cr
& = \underbrace{\lambda q_{\rm o} e^{- \frac{\lambda q_{\rm o}}{M}}}_{=\uE[S]} \frac{1}{B_{\rm o} \mu_{\rm max} } \uE\left[ \frac{1}{\log_2 (1+ \tilde \gamma_{(m)})} 
\right] \cr 
& \le \lambda q_{\rm o} e^{- \frac{\lambda q_{\rm o}}{M}}
\frac{\ln 2}{B_{\rm o} \mu_{\rm max}} \left(\frac{1}{2} + \uE\left[\frac{1}{ \tilde \gamma_{(m)}} \right]
\right) \label{EQ:L1} \\    
& \le \lambda q_{\rm o} e^{- \frac{\lambda q_{\rm o}}{M}}
\frac{\ln 2}{B_{\rm o} \mu} \left(\frac{1}{2} + \uE\left[\frac{1}{ \tilde \gamma_{(m)}} \right] \right), \label{EQ:L1_b}
\end{align}
where 
\be 
\mu_{\rm max} = 
\frac{1}{\uE[\tilde U_{(m)}\,|\, \tilde U_{(m)} \le U_{\rm max} ]} = \frac{\mu (1 - e^{-\mu U_{\rm max}})}
{1 - e^{-\mu U_{\rm max}} (1+ \mu U_{\rm max})}.
\ee 
\end{mylemma}
As shown in \eqref{EQ:L1}, $\uE \left[\frac{1}{\tilde \gamma_{(m)} } \right] < \infty$ of the assumption of {\bf A3} is a sufficient condition for the existence of a finite $\uE\left[ \frac{1}{\log_2 (1+ \tilde \gamma_{(m)})} \right]$.
\begin{IEEEproof}
See Appendix~\ref{A:1}.
\end{IEEEproof} 

Note that we can have the following closed-form expression for $\uE\left[\frac{1}{ \tilde \gamma_{(m)}} \right]$ if the distribution of $\tilde g_{(m)}$ is given as in \eqref{EQ:gdist}:
\begin{align}
\uE\left[\frac{1}{ \tilde \gamma_{(m)}} \right]
= \frac{1}{\bar \gamma} e^{\frac{\Gamma}{\bar \gamma}}
E_1 \left( \frac{\Gamma}{\bar \gamma} \right), 
\end{align} 
where $E_1 (x) = \int_x^\infty \frac{e^{-z}}{z} d z$ is the exponential integral and $\Gamma = \Gamma_{(k)}$ for all $k$.

Recall that active devices are to send the offloading requests through RAC in stage 1. Since there are $M$ channels, multiple active devices can successfully send their offloading requests at a time. As shown in \eqref{EQ:BBB}, as $M$ increases, the uploading time increases (because $B_{\rm o}$ decreases as shown in \eqref{EQ:T_n}). In addition, the increase of $U_{\rm max}$ results in more offloading devices and the increase of the uploading time. Formally, we have the following observations.

\begin{mylemma} \label{L:2}
The mean of $D$, $\uE[D]$, increases \emph{i)} with $M$ and \emph{ii)} with $U_{\rm max}$ for $\lambda \le M$.
\end{mylemma}
\begin{IEEEproof}
For the first part, in \eqref{EQ:L1}, we can see that $\uE[S]$ increases with $M$.
In addition, we have
$\uE[\tilde T_{(m)}] \propto \frac{1}{B_{\rm o}} = 
\frac{1}{B - M b}$ from \eqref{EQ:BBB}. Thus, $\uE[D]$ increases with
$M$.

For the second part, since $\uE[ \tilde U_{(m)} \,|\, \tilde U_{(m)} \le U_{\rm max}]$, it increases with $U_{\rm max}$. In addition, from \eqref{EQ:qo}, we can also see that $q_{\rm o}$ increases with $U_{\rm max}$. Thus, if $\lambda \le M$, $\uE[S] = \lambda q_{\rm o} e^{- \frac{\lambda q_{\rm o}}{M}}$ increases with $q_{\rm o}$ or $U_{\rm max}$. Since $\uE[S]$ and $\uE[ \tilde U_{(m)} \,|\, \tilde U_{(m)} \le U_{\rm max}]$ increase with $U_{\rm max}$, $\uE[D]$ increases with $U_{\rm max}$,
which completes the proof.
\end{IEEEproof}

In addition, suppose that each offloading device can perform ideal power control so that the SNR can be constant, i.e., $\gamma_{(k)} = \bar \gamma > 0$. 
Then, from \eqref{EQ:L1}, we can show that 
\begin{align} 
\mu_{\rm max} \uE[D]
& = \frac{\uE[S]  }{B_{\rm o} \log_2 (1+ \bar \gamma)} \cr  
& \le \frac{M e^{-1}}{B_{\rm o} \log_2 (1+ \bar \gamma)},
\end{align} 
since $x e^{-x} \le e^{-1}$ for $x\ge 0$.
From this, 
a sufficient condition for the stable system becomes
\be 
\frac{M e^{-1}}{ \mu_{\rm max} } \le
\underbrace{\Delta B_{\rm o} \log_2 (1+ \bar \gamma) }_{= \ {\rm number\ of\ Bits\ per\ Round}} ,
    \label{EQ:SC}
\ee 
where $\frac{e^{-1}M}{ \mu_{\rm max} } $ is the product of the maximum average number of offloading devices, $\max \uE[S]= M e^{-1}$, and the average size of input data, $\uE[ \tilde U_{(m)} \,|\, \tilde U_{(m)} \le U_{\rm max}] = \frac{1}{\mu_{\rm max}}$. 
That is, the left-hand side (LHS) term on \eqref{EQ:SC} is the average number of input bits per round, while the right-hand side (RHS) term is the average number of bits that can be transmitted through OC per round.

Consequently,  there are two key control parameters, $M$ and $U_{\rm max}$ in the two-stage approach. While $U_{\rm max}$ is used for local decision of offloading at devices, $M$ is a system parameter that can be decided to limit the number of offloading devices, $S$, so that the total upload time does not grow for a stable system (e.g., if $M$ decreases, the total upload time decreases because the number of offloading devices, $S$, decreases as well as the bandwidth for uploading, $B_{\rm o}$, increases).
That is,
to stabilize the system, the value of $M$ or $U_{\rm max}$ can be controlled. For example, from Lemma~\ref{L:2}, $U_{\rm max}$ can be adjusted as follows:
\be 
\hat U_{\rm max} (i+1) = \hat U_{\rm max} (i) - \eta_i ( D_{i}-\Delta),
    \label{EQ:SA}
\ee 
where $\eta_i > 0$ is the step size and $\hat U_{\rm max} (i)$ stands for the value of $U_{\rm max}$ used in round $i$. Provided that 
$$
\lim_{U_{\rm max} \to \infty} \uE[D] > \Delta,
$$
$\hat U_{\rm max}(i)$ in \eqref{EQ:SA} can approach the value of $U_{\rm max}$ that satisfies $\uE[D] = \Delta$. In particular, as shown in Lemma~\ref{L:2}, $\uE[D]$ is a nondecreasing function of $U_{\rm max}$. Thus, 
if $D_i$ is iid, with a sequence of $\eta_i$ satisfying $\sum_i \eta_i = \infty$ and $\sum_i \eta_i^2 < \infty$, $\hat U_{\rm max} (i)$ in \eqref{EQ:SA} converges to the solution of the equation $\uE[D] = \Delta$ with probability 1 \cite{Kushner03}.
Likewise,  the value of $M$ can be adaptively decided to keep the system stable. 

%At the end of each round, $D_i$, can be sent back to all devices so that they can adjust $U_{\rm max}$ or $M$. 
%The resulting approaches are referred to as the stability-based approaches.

It is noteworthy that the devices are discouraged to offload their tasks as $U_{\rm max}$ and $M$ decrease, while decreasing $M$ and $U_{\rm max}$ have different implications. The decrease of $M$ is independent of the offloading probability. Thus, when $M$ decreases, more contending devices experience collisions in stage 1, which are then forced to perform local computing (accordingly, such devices have a disadvantage in that local computing is delayed by $\Delta$), 
while the total upload time effectively decreases as the bandwidth of OC increases (see \eqref{EQ:BBB}) as well as the number of offloading devices decreases. 
On the other hand, a decrease in $U_{\rm max}$ can directly reduce the number of contending devices, thereby reducing the burden on stage 1 and saving the decision time for local computing. However, adjusting the value of $M$ provides a  better control over the total upload time, $D_i$, compared to $U_{\rm max}$ (related simulation results will be presented in Section~\ref{S:Sim}.

Unfortunately, while a finite offloading latency can be obtained (by ensuring a stable system), it is difficult to guarantee a specific offloading latency. Thus, in the next section, we consider the latency outage from a device perspective to see if a certain offloading latency target can be met.

\section{Device Perspective: Analysis of Latency Outage}    \label{S:LO}

In this section, we analyze its latency outage probability
from a device perspective. For simplicity, we assume that $U_{\rm max} = \infty$ (i.e., all the active devices become the contending devices so that $W = K$) unless stated otherwise.

%the case that all active devices become contending devices, i.e., $q_{\rm o} = 1$ and $K = W$, and derive 

\subsection{Latency Outage Probability} \label{SS:LOP}

Suppose that active
device $k$ can have
the following latency threshold that increases with its upload time:
\be 
\tau_{(k)} = T_{(k)} +\tau, 
    \label{EQ:tau}
\ee
where $\tau > 0$ (which 
is an additional latency in addition to its own upload time, $T_{(k)}$, to account for other previously scheduled uploads), and wants to know the probability that the upload can be completed with the latency threshold, $\tau_{(k)}$.
The BS will schedule their uploading according to a certain order, which is unknown to active devices. 
Active device $k$ can assume that it succeeds to send offloading request and its upload is placed the $S(k)$th order among the $S$ active devices that successfully send the requests in stage 1,
where $1 \le S(k) \le S$.
Then, active device $k$ can expect to complete the upload at the following time:
\be 
Y_{(k)} = t_N + \underbrace{ \sum_{m=1}^{S_{(k)} -1} \tilde T_{(m)} }_{=Z_{(k)}} + T_{(k)}.
    \label{EQ:Yk}
\ee 
In \eqref{EQ:Yk}, $t_N$ is known as the BS sends this information prior to the random access round and $T_{(k)}$ is also known to active device $k$. However, $Z_{(k)}$ is unknown as it depends on $S$ and the scheduling order. 
Then, from \eqref{EQ:tau} and \eqref{EQ:Yk},
the probability
that active device $k$ fails to meet the latency constraint
is given by
\begin{align}
\Pr(Y_{(k)} > \tau_{(k)} )
& = \Pr \left( Z_{(k)} > \tau- t_N \right) \cr 
& = g\left( Z_{(k)} > \tau- t_N \right), 
    \label{EQ:Yt}
\end{align} 
where $g(\cdot)$ is the complement of the cumulative distribution function (cdf)
of $Z_{(k)}$, which will be discussed in Subsection~\ref{SS:UB}.
This probability, which will be referred to as the latency outage probability, is a function of $\tau-t_N$. From the known state variable, $X_i = x_N$, at round $i$, each device can find the latency outage probability as a quality-of-service (QoS) indicator.

\subsection{An Upper-bound} \label{SS:UB}

In this subsection, we consider a closed-form expression for an upper-bound on the function $g(\cdot)$ in \eqref{EQ:Yk}.

In \eqref{EQ:Yk}, $Z_{(k)}$ represents the delay due to the uploads of the other active devices at $\tilde t = 0$ scheduled before device $k$, which is referred to as the intra-delay.
Note that 
in \eqref{EQ:Yk}, we simply assume that devices $m= 1, \ldots, S_{(k)}-1$ are the devices successfully sending the requests and their upload orders are placed before device $k$. To device $k$, at time $t$, it is unknown that which are the devices that successfully send their requests and are placed earlier than it for uploading. Thus, $\tilde T_{(m)}$, $m \ne k$, are random variables to active device $k$.

To characterize $Z_{(k)}$ at the $k$th active device, under 
the assumptions of {\bf A1} and {\bf A2},
we can consider the following approximation:
\be
\tilde T_{(m)} \approx 
V_{(m)} = \frac{\tilde U_{(m)}}{B_{\rm o}} 
\uE \left[ \frac{1}{\log_2 (1+  \gamma)} \right]
\sim {\rm Exp} (\theta),
    \label{EQ:tXe}
\ee 
where
\be 
\frac{1}{\theta} = \frac{1}{\mu B_{\rm o}} \uE \left[ \frac{1}{\log_2 (1+  \gamma)} \right]  .
\ee 
Thus, $Z_{(k)}$ can be seen as a sum of $S_{(k)} - 1$ exponential random variables with parameter $\theta$. If the variation of the SNR is not significant, \eqref{EQ:tXe} becomes a good approximation. In particular, if an ideal power control is employed so that $\gamma_{(k)}$ becomes a constant, \eqref{EQ:tXe} becomes accurate.

Using the Chernoff bound \cite{Mitz05}, an upper-bound on  the latency outage probability can be obtained as follows:
\begin{align}
\Pr (Z_{(k)} > \tau) 
& = g(\tau) \cr
& \le \bar g(\tau) = \min_{\nu \ge 0} e^{-\nu \tau} 
\uE[ e^{\nu Z_{(k)}} ] ,
    \label{EQ:CB}
\end{align} 
where the upper-bound, $\bar g(\tau)$, decreases exponentially with $\tau$.
Assuming that the BS randomly schedules the uploads for $S$
offloading devices, $S_{(k)}$ can have the following
distribution:
\be 
\Pr(S(k) = s\,|\, S) = \frac{1}{S}, \ s \in \{ 1,\ldots, S \}.
\ee 
Then, from \eqref{EQ:tXe}, after some manipulations, we can show that
\begin{align}
\uE [ e^{\nu  Z_{(k)}}  ] 
& = \uE \left[ \left(
\frac{\theta}{\theta - \nu} \right)^{S(k)-1}
\right] \cr 
& = \uE \left[\uE \left[ \frac{1}{S} \sum_{s=0}^{S-1}\left(
\frac{\theta}{\theta - \nu} \right)^{s} \,\biggl|\, S
\right] \right] \cr 
& = \uE \left[ \frac{1}{S} \left(\frac{1- z^S}{1-z} \right) \right],
    \label{EQ:EvZ}
\end{align} 
where $z = \frac{\theta}{\theta - \nu}$.
To find a tight upper-bound in \eqref{EQ:CB}, we need to have a closed-form expression for $\uE [ e^{\nu  Z_{(k)}}  ]$. To this end, we consider 
another approximation.

For a sufficient large $M$, the number of offloading devices, $S$, can be approximated by a Poisson random variable with mean $\uE[S] = 
\bar \lambda = \lambda  e^{-\frac{\lambda }{M}}$ (because we consider the case of $q_{\rm o}=1$).
Based on this Poisson approximation, we have the following result.

\begin{mylemma} \label{L:3}
If $S$ is a Poisson random variable and $\tilde T_{(m)}$ is an 
exponential random variable as in \eqref{EQ:tXe}, we have
\begin{align}
\uE [ e^{\nu  Z_{(k)}}  ] = 
\frac{e^{-\bar \lambda}}{1-z} \lim_{n_{\rm max} \to \infty}
\sum_{n=2}^{n_{\rm max}} \bar \lambda \beta_n (\bar \lambda) - 
 \bar \lambda z \beta_n (\bar \lambda z),
    \label{EQ:L3}
\end{align}
where
\be 
\beta_n (x) = 
\frac{(n-2)!}{x^n} \left( e^x - \sum_{s=0}^{n-1} \frac{x^s}{s!} \right) .
\ee 
Here, with a finite $n_{\rm max} \ge  2$,  we can have an approximation of $\uE [ e^{\nu  Z_{(k)}}]$.
\end{mylemma}
\begin{IEEEproof}
See Appendix~\ref{A:3}.
\end{IEEEproof}

With a sufficiently large $n_{\rm max}$, we can obtain
$\uE [ e^{\nu  Z_{(k)}}  ]$ in \eqref{EQ:L3}, which can be used for the minimization in \eqref{EQ:CB} for the Chernoff-bound.
Note that if $U_{\rm max}$ is finite, $q_{\rm o}$ can be less than 1. In this case, with $\bar \lambda =\lambda q_{\rm o} e^{-\frac{\lambda q_{\rm o}}{M}}$, we can also obtain $\uE [ e^{\nu  Z_{(k)}}  ]$ from \eqref{EQ:L3}.
\section{Numerical Results} \label{S:Sim}

In this section, we present theoretical and simulation results under the assumptions of {\bf A1} -- {\bf A3} with \eqref{EQ:gdist}. For convenience, the bandwidth of one channel for RAC, $b$, is normalized (i.e., $b = 1$), and $B = b M_{\rm max} = M_{\rm max}$, where $M_{\rm max}$ represents the maximum number of channels for RAC.

\subsection{System Perspective}

\begin{figure}[thb]
\begin{center} 
\subfigure[Performance in terms of $\lambda$]{\label{fig 0 ay}
\includegraphics[width=0.4\textwidth]{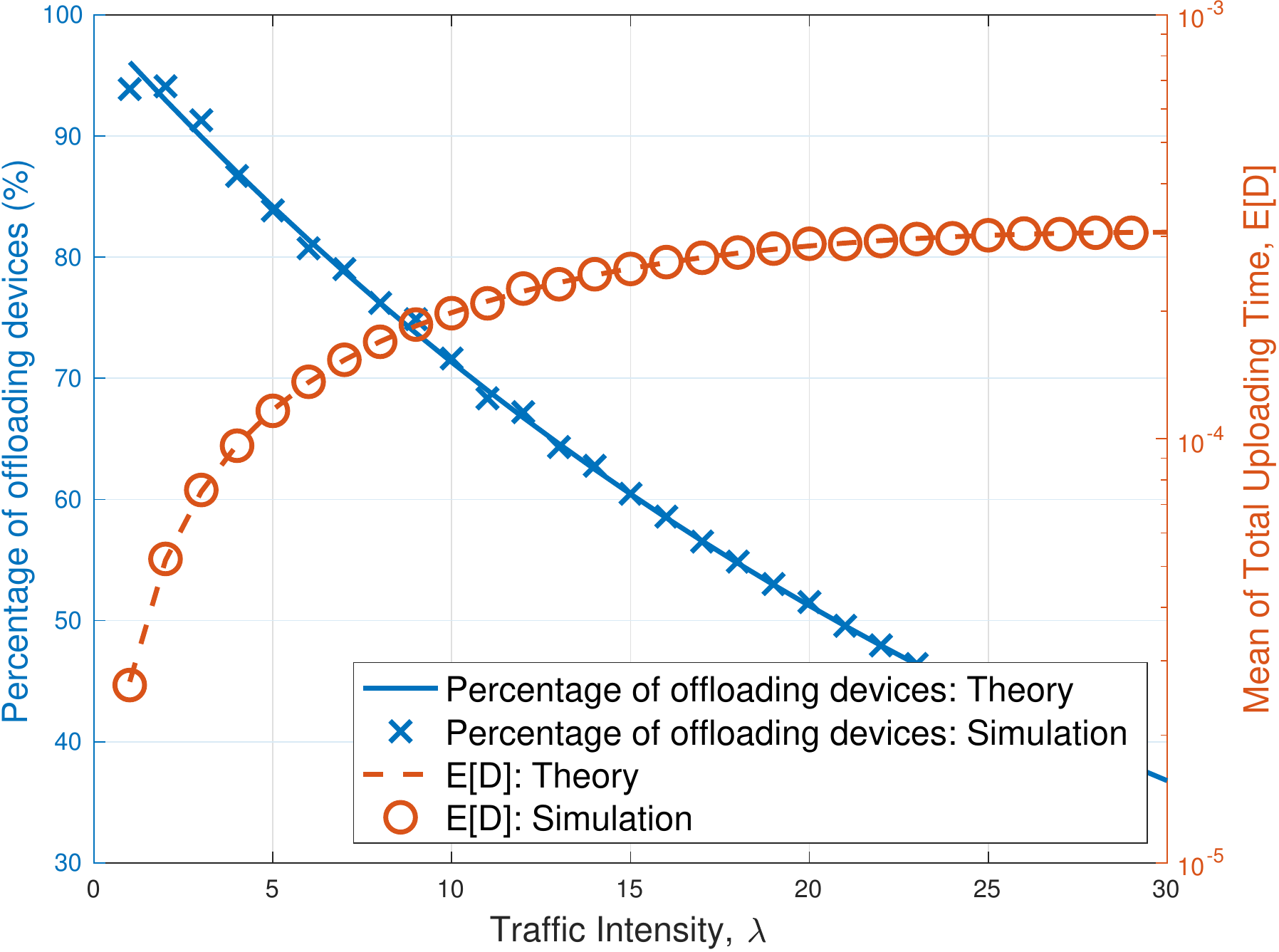}}
\subfigure[Performance in terms of $\mu$]{\label{fig 0 by}
\includegraphics[width=0.4\textwidth]{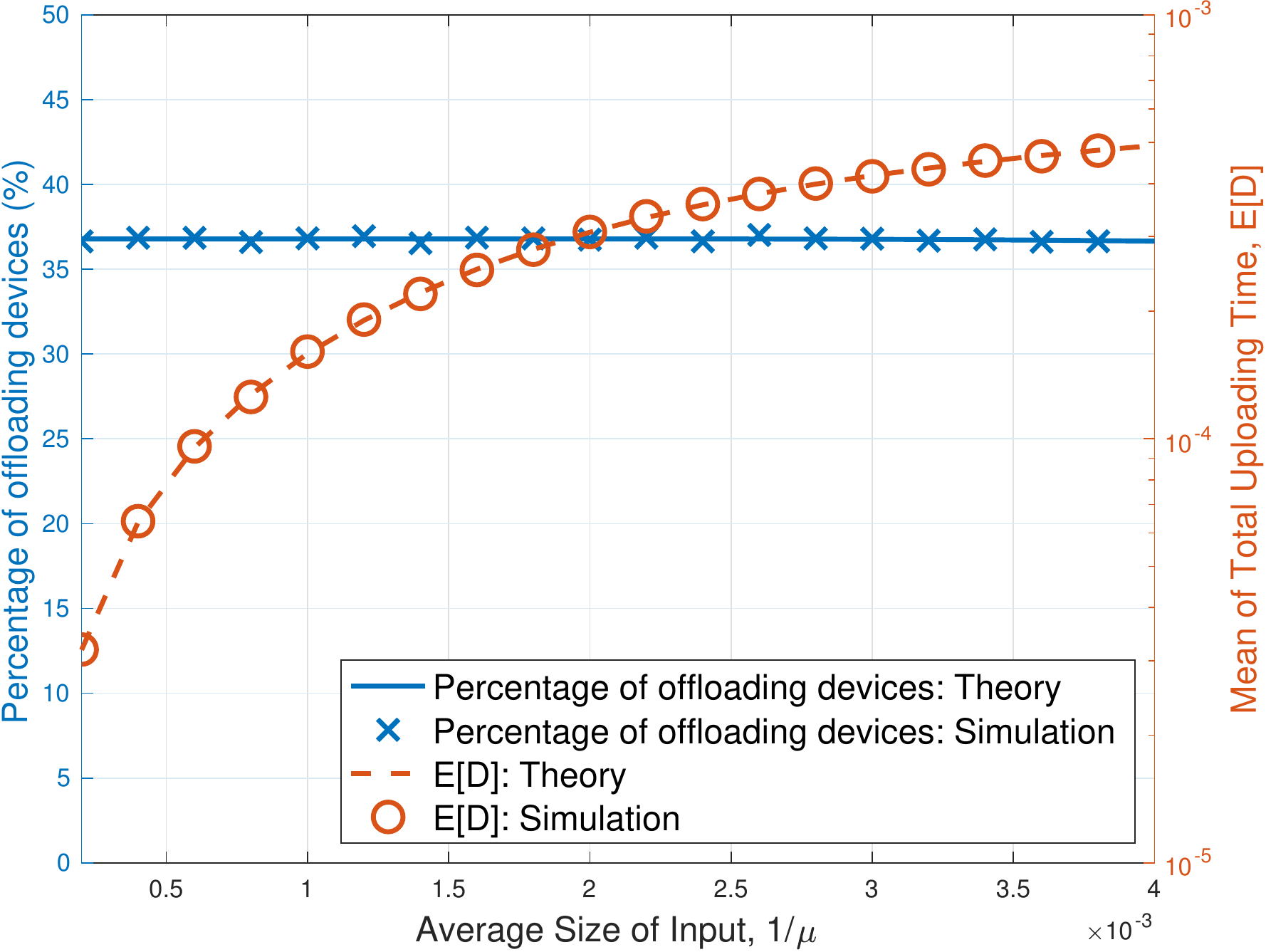}} %\vspace{-1em}
\end{center}
\caption{Percentage offloading devices and total upload time
when $M_{\rm max} = B = 50$, $M = 30$, $\Delta = 10^{-3}$, $U_{\rm max} = 10 \Delta$, 
and $(\bar \gamma, \Gamma_{(k)}) = (10, 6)$ in dB: (a) for different values of $\lambda$ with $\frac{1}{\mu} = 2 \Delta$; (b) for different values of $\mu$ with $\lambda = 30$.}
        \label{Fig:plt1}
\end{figure}

In Fig.~\ref{Fig:plt1}, we show the performance in terms of 
the percentage of offloading devices, $\frac{\uE[S]}{\lambda}$ in \%, and the average total upload time, $\uE[D]$,
for different values of the traffic intensity or average number of new devices per round, $\uE[K] = \lambda$, and the average size of input, $\uE[U_{(k)}]= \frac{1}{\mu}$, 
when $M_{\rm max} = B = 50$, $M = 30$, $\Delta = 10^{-3}$, $U_{\rm max} = 10 \Delta$, 
and $(\bar \gamma, \Gamma_{(k)}) = (10, 6)$ in dB. We can see that most devices can offload
when the system is lightly loaded (i.e., $\lambda$ is small) in Fig.~\ref{Fig:plt1} (a). As $\lambda$ approaches $M$, the system becomes fully loaded and about $36\%$ of active devices can offload their tasks as shown in Fig.~\ref{Fig:plt1} (a) and (b) (as $\uE[S] \le M e^{-1}$).
It is also shown that $\uE[D]$ increases with $\lambda$ and $\frac{1}{\mu}$, while $\uE[D]$ is less than $\Delta = 10^{-3}$. This shows that the system is stable (i.e., $\uE[D] \le \Delta$). In fact, with $M = 30$ and $\frac{1}{\mu} = U_{\rm max} = 10 \Delta$, from \eqref{EQ:L1}, we have $\uE[D] = 6.06 \times 10^{-4} < \Delta = 10^{-3}$, which indicates that the system is stable for all the range of parameter values in Fig.~\ref{Fig:plt1} (a) and (b), where we can also find that the theoretical results agree with the simulation results.

%In Fig.~\ref{Fig:plt1} (a), we note that the increase of $\lambda$ results in the increase of $\uE[S]$ and $\uE[D]$. 

\begin{figure}[thb]
\begin{center} 
\subfigure[Performance in terms of $M$]{\label{fig 1 ay}
\includegraphics[width=0.4\textwidth]{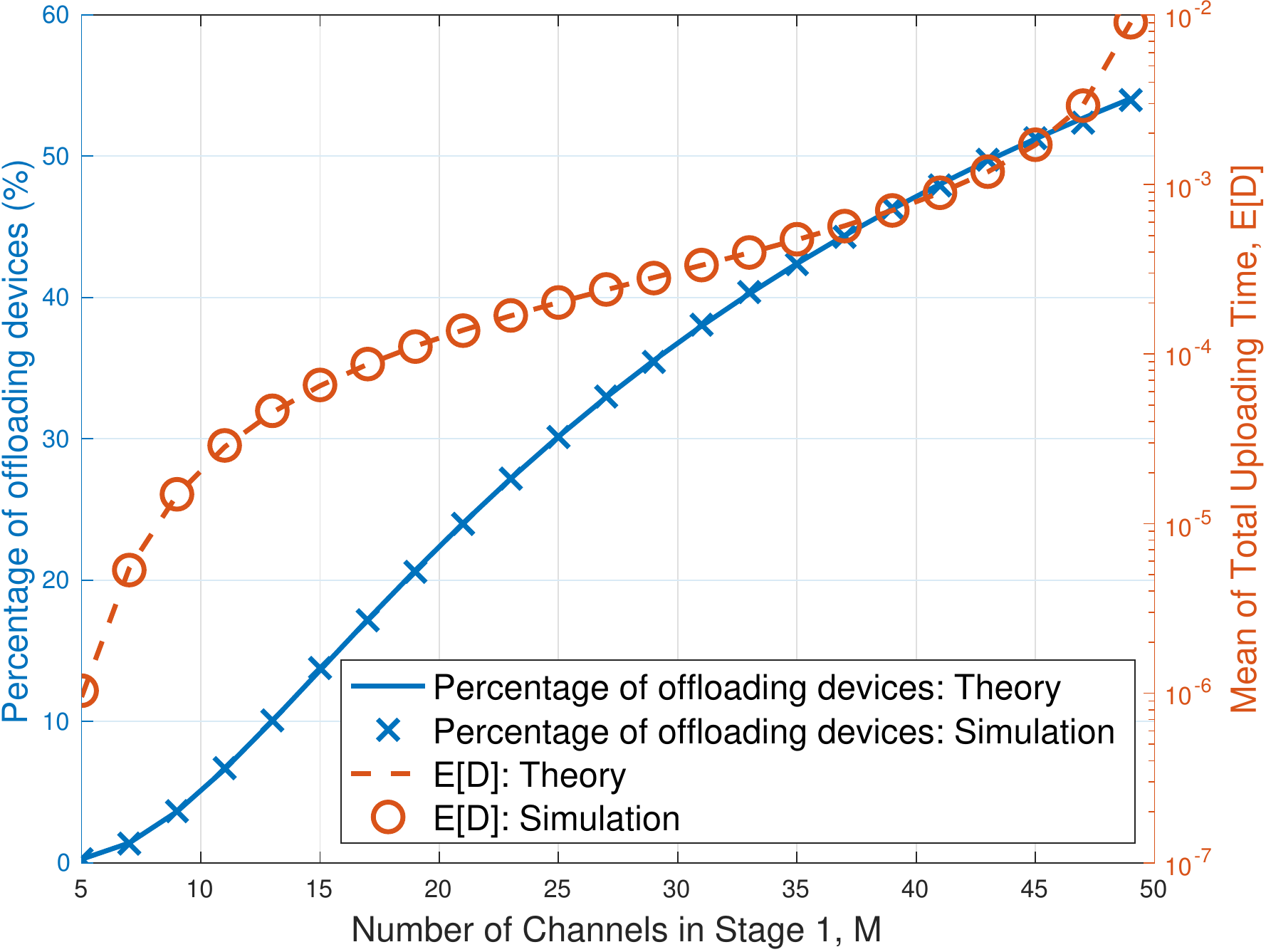}}
\subfigure[Performance in terms of $U_{\rm max}$]{\label{fig 1 by}
\includegraphics[width=0.4\textwidth]{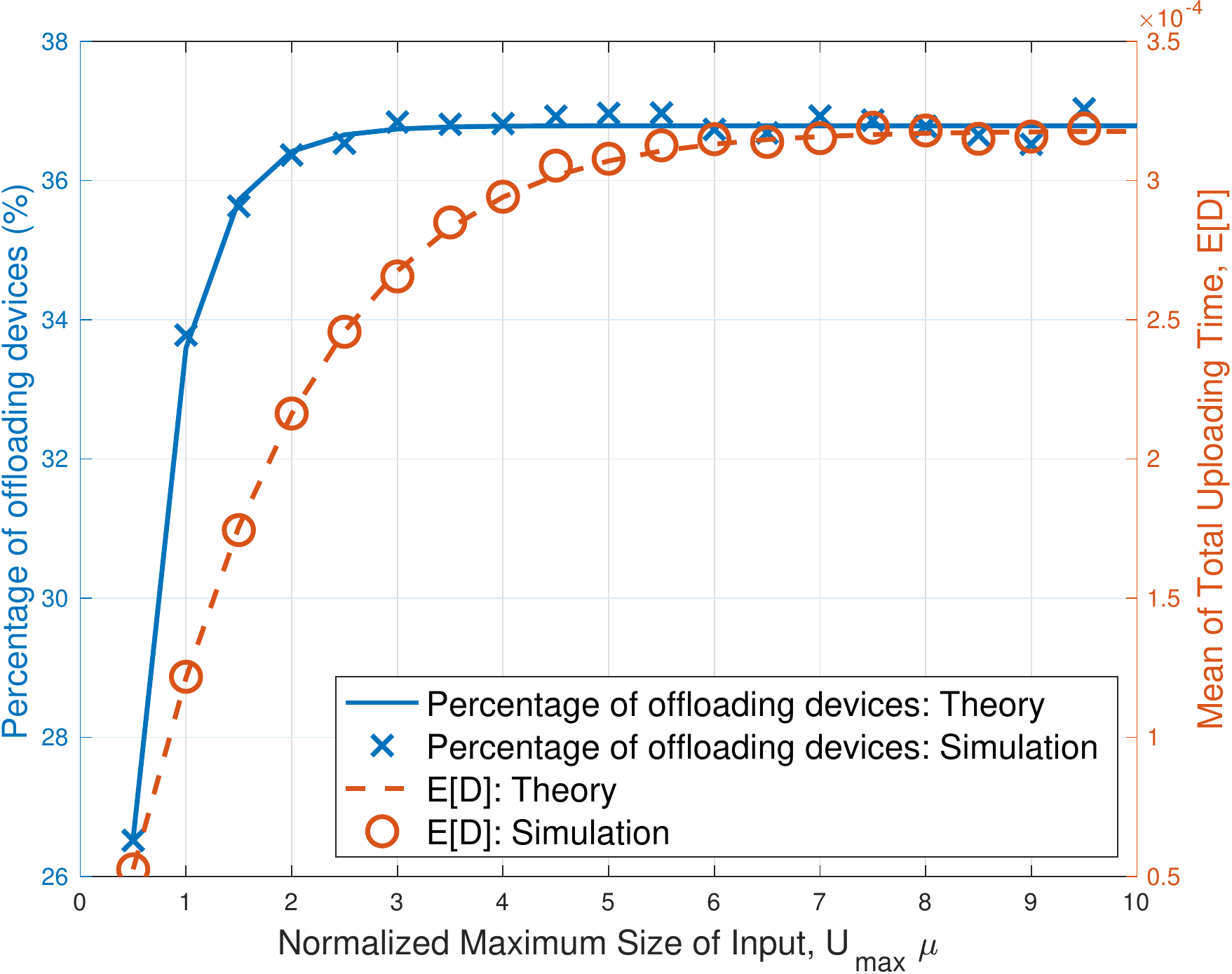}} %\vspace{-1em}
\end{center}
\caption{Percentage offloading devices and total upload time
when $M_{\rm max} = B = 50$, $\Delta = 10^{-3}$, $\frac{1}{\mu} = 2 \Delta$, $\lambda = 30$,
and $(\bar \gamma, \Gamma_{(k)}) = (10, 6)$ in dB: (a) for different values of $M$ with $U_{\rm max} = 10 \Delta$; (b) for different values of $U_{\rm max}$ with $M = 30$.}
        \label{Fig:plt2}
\end{figure}

In Fig.~\ref{Fig:plt2}, $\lambda$ and $\mu$ are fixed, while 
$M$ and $U_{\rm max}$ vary when $M_{\rm max} = B = 50$, $\Delta = 10^{-3}$, $\frac{1}{\mu} = 2 \Delta$, $\lambda = 30$,
and $(\bar \gamma, \Gamma_{(k)}) = (10, 6)$ in dB. We can see that $\uE[D]$ becomes greater than $\Delta$ when $M \ge 40$ in Fig.~\ref{Fig:plt2} (a). Clearly, since more devices choose to offload as $M$ increases, the total upload time increases and the system can be unstable. On the other hand, in Fig.~\ref{Fig:plt2} (b), we see that although $U_{\rm max} \to \infty$, $\uE[D]$ is still less than $\Delta$, which results from the fact that $\uE[D] \to 3.17 \times 10^{-4}$ as $\mu_{\rm max} \to \mu$ (which is the case when $U_{\rm max} \to \infty$). As mentioned earlier, 
when comparing Fig.~\ref{Fig:plt2} (a) and (b), it can be seen that compared to $U_{\rm max}$, $M$ is a system parameter to more effectively control the mean of the total upload time, $\uE[D]$.

\begin{figure}[thb]
\begin{center} 
\subfigure[Adaptation of $M$]{\label{fig 2 ay}
\includegraphics[width=0.4\textwidth]{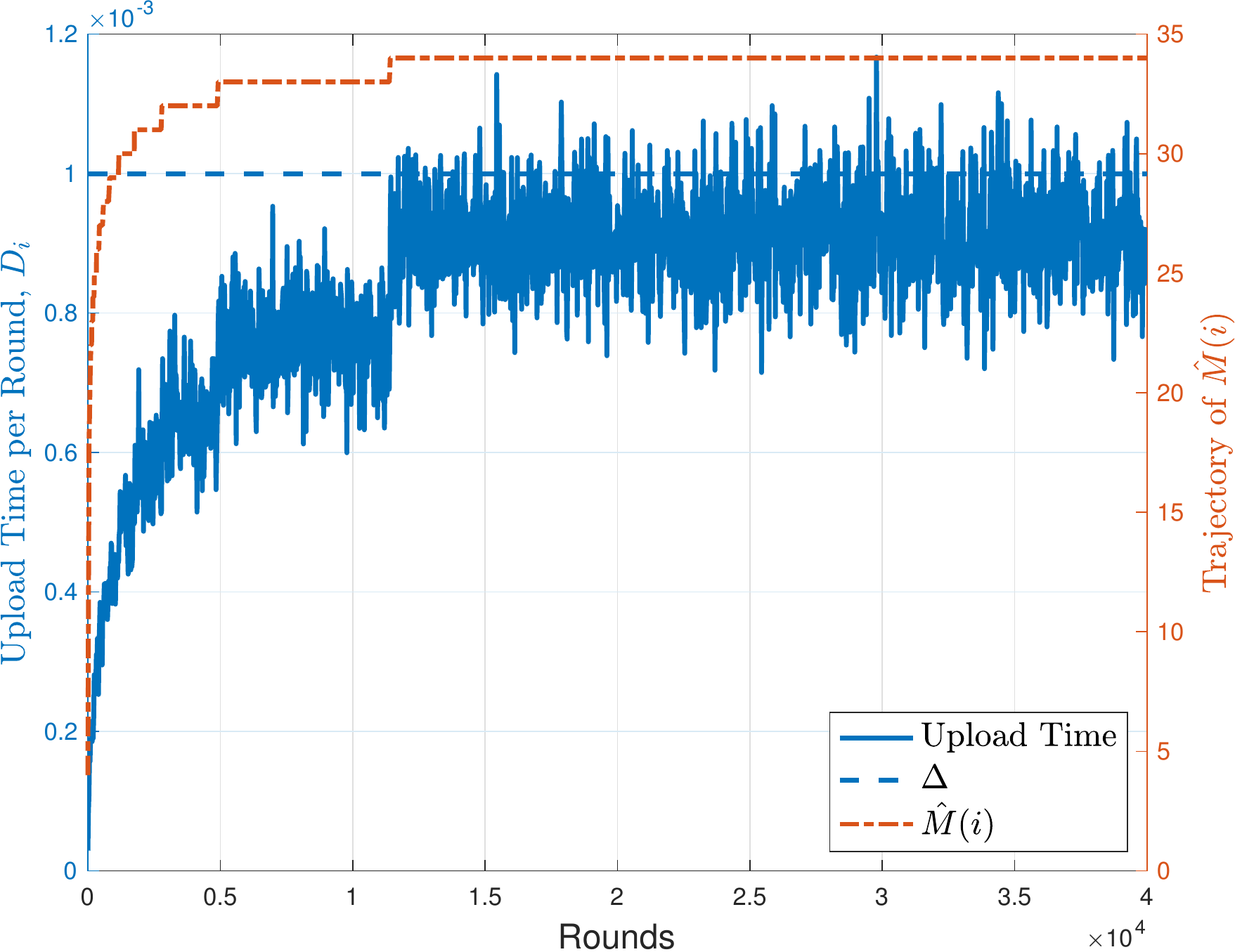}}
\subfigure[Adaptation of $U_{\rm max}$]{\label{fig 2 by}
\includegraphics[width=0.4\textwidth]{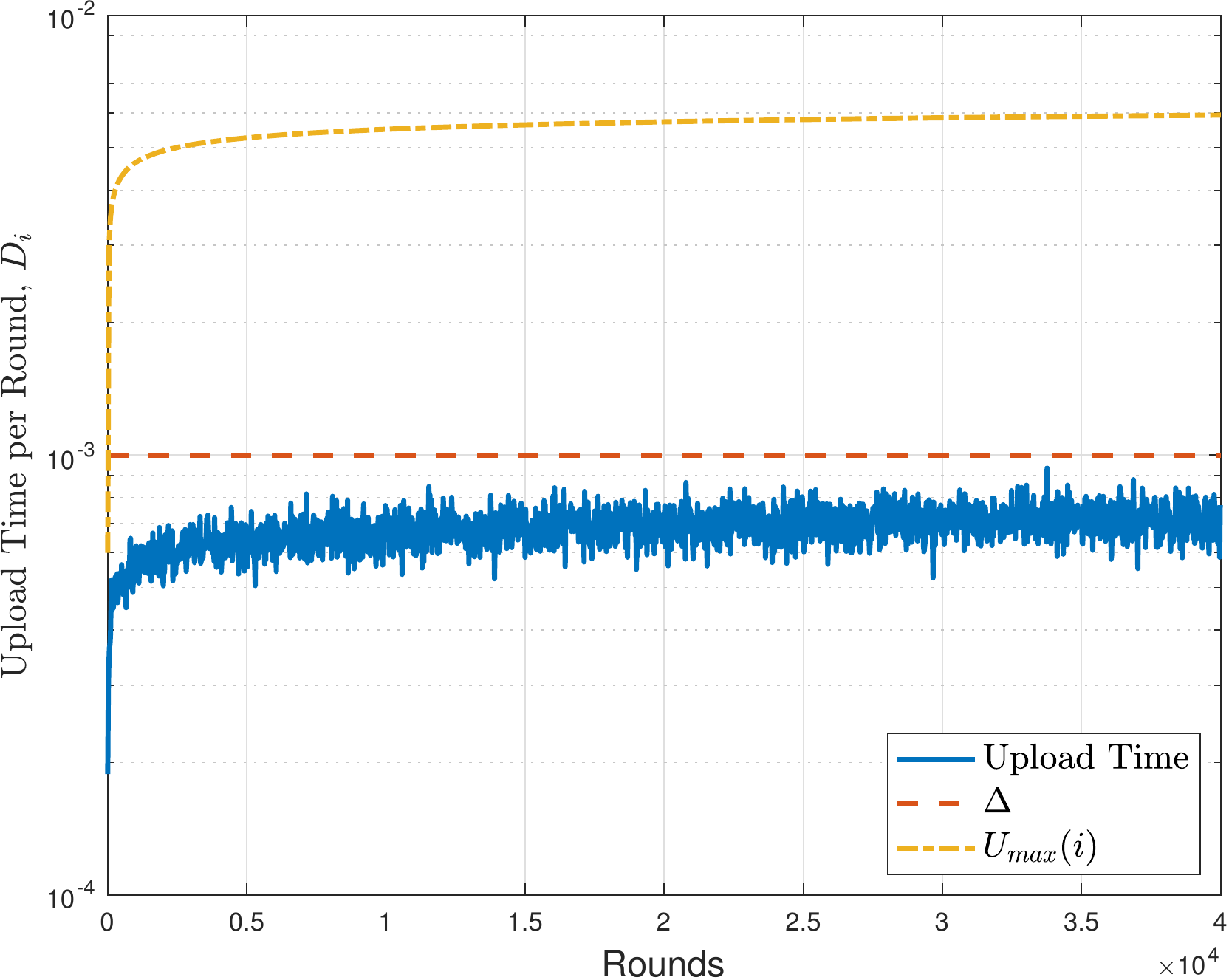}} %\vspace{-1em}
\end{center}
\caption{Total upload time and trajectory of key parameters over time: (a) $\hat M(i)$ with $\frac{1}{\mu} = 2 \Delta$ and $U_{\rm max} = 5 \Delta$; (b) $\hat U_{\rm max}(i)$ with $\frac{1}{\mu} = 5 \Delta$ and $M = 30$.}
        \label{Fig:plt3}
\end{figure}

As mentioned earlier, it is possible to adjust the values of $M$ and/or $U_{\rm max}$ to ensure $\uE[D] = \Delta$ using the iterative method for stochastic approximation in \eqref{EQ:SA}.
In Fig.~\ref{Fig:plt3}, we show the results when 
$M_{\rm max} = B = 40$, $\Delta = 10^{-3}$,  $\lambda = 30$,
and $(\bar \gamma, \Gamma_{(k)}) = (10, 6)$ in dB. For the case that $M$ is adapted as in Fig.~\ref{Fig:plt3} (a), we assume that $\frac{1}{\mu} = 2 \Delta$ and $U_{\rm max} = 5 \Delta$. It is shown that $M (i) \to 34$ as the number of rounds, $i$, increases, while $D_i$ is around $\Delta$. 
As shown in Fig.~\ref{Fig:plt3} (b), $U_{\rm max}$ can also be adjusted, where $U_{\rm max}(i) \to 5.94 \times 10^{-3}$ as $i$ increases. Note that $\uE[D] \to 1.58 \times 10^{-3}$ as $U_{\rm max} \to \infty$ in the case of Fig.~\ref{Fig:plt3} (b). Thus, there exists $U_{\rm max}$ that satisfies $\uE[D] = \Delta$. As shown in Fig.~\ref{Fig:plt3} (a) and (b),
we can see that the adaptation of $M$ and $U_{\rm max}$ can make the system stable using an iterative method for stochastic approximation.

\subsection{Latency Outage Probability} 
In this subsection, we focus on the latency outage probability with
$U_{\rm max} = \infty$ (unless stated otherwise) and $\tau_N = 0$. In addition, $n_{\rm max}$ in \eqref{EQ:L3} is set to 20 in finding the upper-bound using \eqref{EQ:CB}.

\begin{figure}[thb]
\begin{center}
\includegraphics[width=\figwidth]{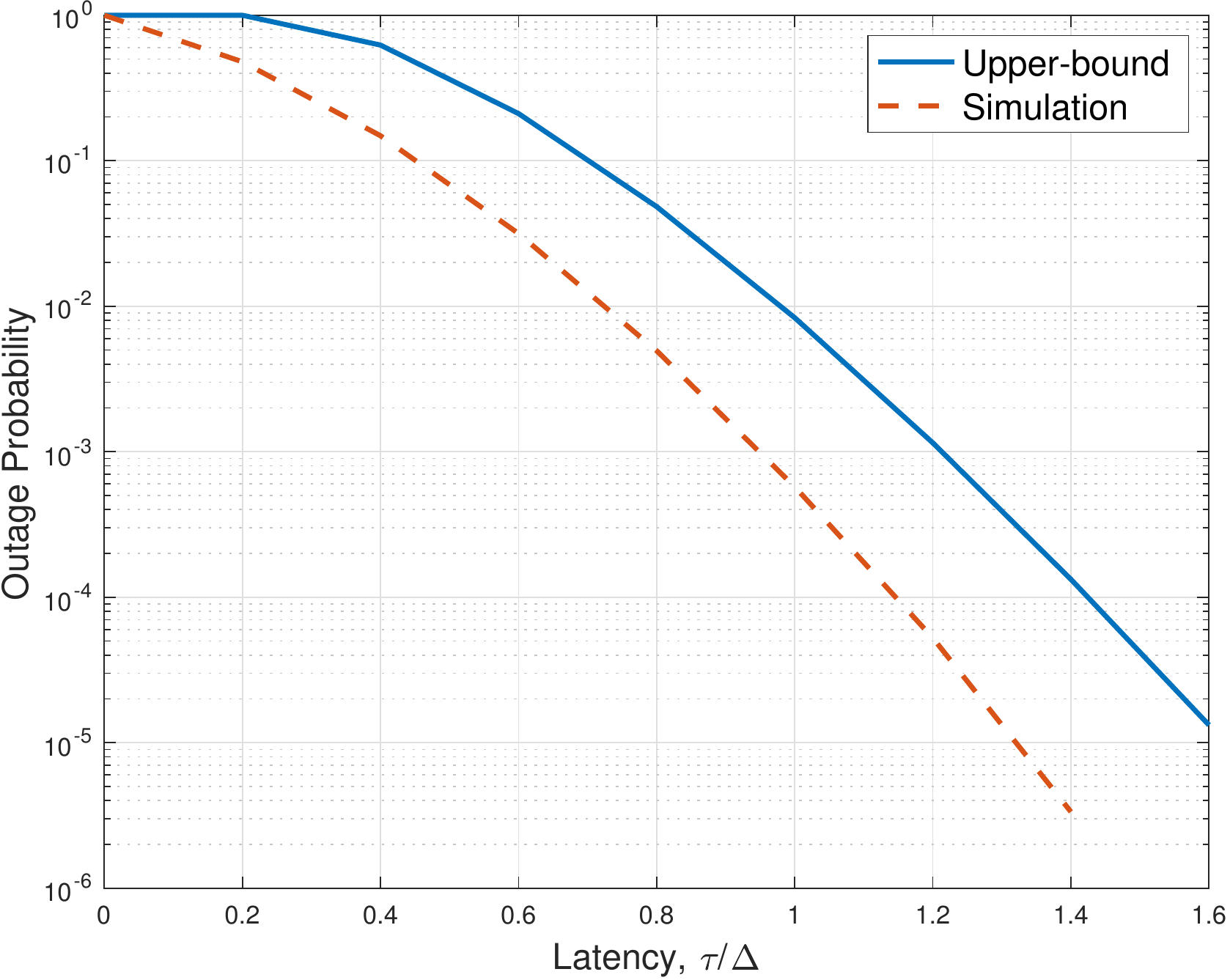} 
\end{center}
\caption{The latency outage probability as a function of $\tau$
when $M_{\rm max} = B = 50$, $\Delta = 10^{-3}$, $\frac{1}{\mu} = 3 \Delta$, $\lambda = 20$, $M = 30$,
and $(\bar \gamma, \Gamma_{(k)}) = (10, 6)$ in dB.}
        \label{Fig:ub1}
\end{figure}

In Fig.~\ref{Fig:ub1}, the latency outage probability is shown as a function of $\tau$
when $M_{\rm max} = B = 50$, $\Delta = 10^{-3}$, $\frac{1}{\mu} = 3 \Delta$, $\lambda = 20$, $M = 30$,
and $(\bar \gamma, \Gamma_{(k)}) = (10, 6)$ in dB. As expected, 
the latency outage probability exponentially decreases with $\tau$. We can also see that \eqref{EQ:CB} is an upper-bound on the latency outage probability, while it is not tight. At a reasonably low outage probability, say $10^{-2} \sim   10^{-4}$, it can be seen that the actual outage probability is about $1/10$ of the upper-bound. Since the Chernoff bound in \eqref{EQ:CB} is known to be asymptotically tight, a scaling factor can be introduced for a good prediction, i.e., $g(\tau) \approx c_0 \bar g(\tau)$, where $c_0$ is a constant. In our case, $c_0$ is around $1/10$.

\begin{figure}[thb]
\begin{center}
\includegraphics[width=\figwidth]{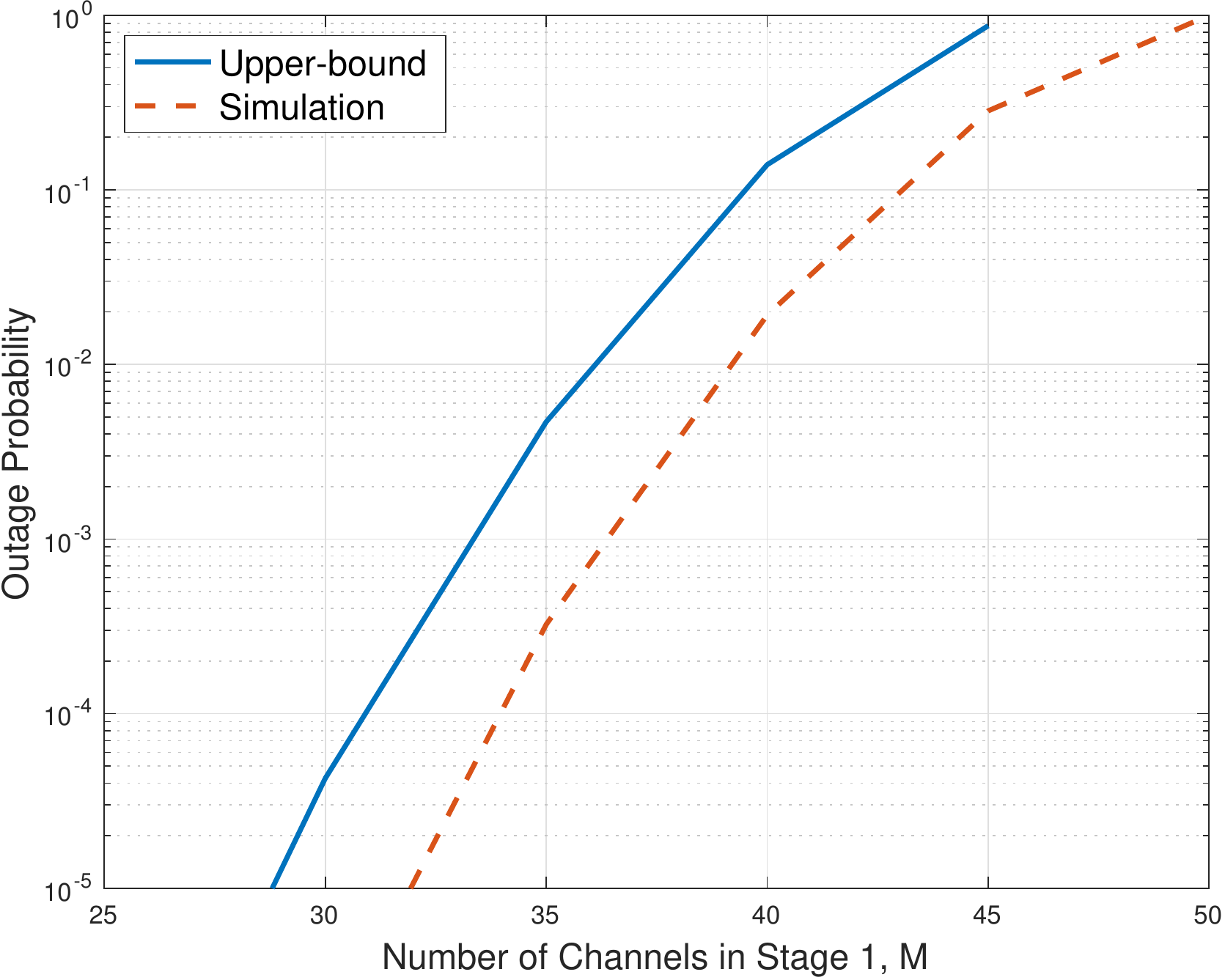} 
\end{center}
\caption{The latency outage probability as a function of $M$
when $M_{\rm max} = B = 50$, $\Delta = \tau = 10^{-3}$, $\frac{1}{\mu} = 2 \Delta$, $\lambda = 20$, 
and $(\bar \gamma, \Gamma_{(k)}) = (10, 6)$ in dB.}
        \label{Fig:ub2}
\end{figure}

Fig.~\ref{Fig:ub2} shows the latency outage probability as a function of $M$
when $M_{\rm max} = B = 50$, $\Delta = \tau = 10^{-3}$, $\frac{1}{\mu} = 2 \Delta$, $\lambda = 20$, 
and $(\bar \gamma, \Gamma_{(k)}) = (10, 6)$ in dB. The latency outage probability decreases as $M$ decreases. Since the decrease of $M$ leads to the increases of the bandwidth of OC, $B_{\rm o}$, and the decrease of the number of the offloading devices, $S$, we can see that the latency outage probability can be significantly lowered by decreasing $M$. Thus, in order to ensure a low latency outage probability, the number of channels for RAC, $M$, should be limited.

\begin{figure}[thb]
\begin{center}
\includegraphics[width=\figwidth]{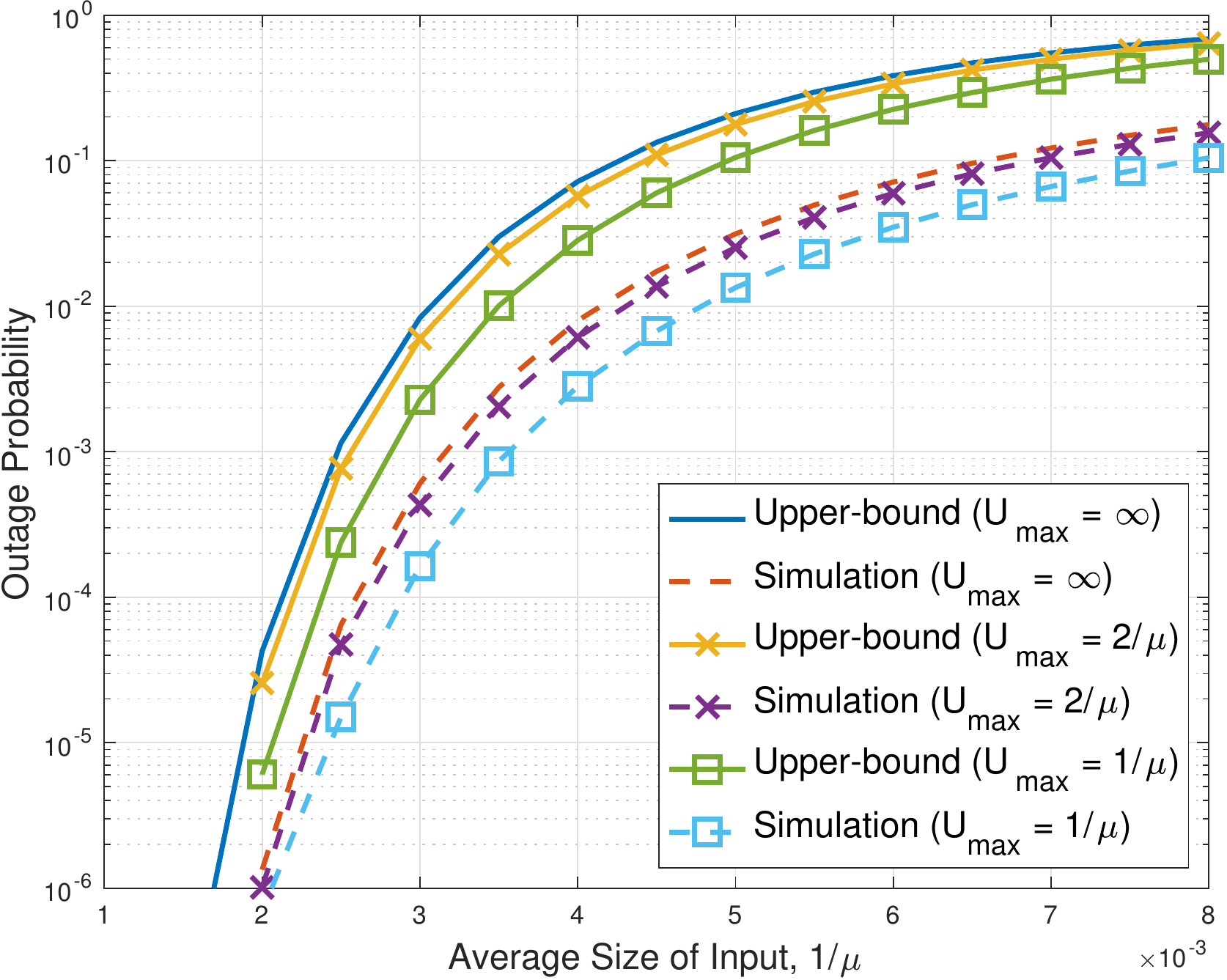} 
\end{center}
\caption{The latency outage probability as a function of $\frac{1}{\mu}$
when $M_{\rm max} = B = 50$, $\Delta = \tau = 10^{-3}$, $\lambda = 20$, $M = 30$,
and $(\bar \gamma, \Gamma_{(k)}) = (10, 6)$ in dB.}
        \label{Fig:ub3}
\end{figure}

The impact of the average size of input, $\frac{1}{\mu}$, on the latency outage probability is shown in Fig.~\ref{Fig:ub3}
when $M_{\rm max} = B = 50$, $\Delta = \tau = 10^{-3}$, $\lambda = 20$, $M = 30$,
and $(\bar \gamma, \Gamma_{(k)}) = (10, 6)$ in dB. It is shown that the increase of $\frac{1}{\mu}$ results in a rapid increase of the 
latency outage probability. By limiting the size of input data using $U_{\rm max}$, a lower latency outage probability can be achieved 
as shown in Fig.~\ref{Fig:ub3}. It is also shown in Fig.~\ref{Fig:plt2} (b) that the decrease of $U_{\rm max}$ leads to the decrease of $\uE[D]$. Thus, the control of $U_{\rm max}$ can be seen as a self-censoring mechanism where a small value of $U_{\rm max}$ discourages offloading. In other words, devices compares their sizes of input to $U_{\rm max}$ and decide offloading themselves
if the system is within a stabilization range or a low enough latency is possible.

\section{Concluding Remarks}    \label{S:Con}

In this paper, we studied multiuser MEC offloading for devices with sporadic tasks in IoT applications. To support offloading of sporadic tasks with low signaling overhead, multichannel random access was employed for offloading requests in the proposed two-stage approach. The two key parameters, the number of channels  for RACs, $M$, and maximum size of input for offloading, $U_{\rm max}$, have been identified to stabilize the system using stochastic approximation, where their values can be adaptively adjusted. Although a finite upload time is guaranteed in a stable system, each device may want to see a QoS indicator. To this end, we analyzed the latency outage probability and its upper-bound was found.

The approach in this paper can be extended in a number of directions. For example, NOMA can be used for stage 2 so that the upload time can be shortened. The notion of priority queue can be introduced to provide different QoS for devices with different latency constraints.

\appendices
\section{Proof of Lemma~\ref{L:1}}  \label{A:1}

In \eqref{EQ:DX}, since $S$ and $\tilde T_{(m)}$ are independent, 
due to Wald's identity \cite{Mitz05}, we have
$\uE[D]  = \uE[S] \uE[\tilde T_{(m)}]$, which is the equality in \eqref{EQ:L1}. 
From \eqref{EQ:WP}, it can be shown that 
\begin{align}
\uE[S] & = \uE\left[\uE[S\,|\, W] \right]  \cr 
& = \uE \left[ W  \left(1 - \frac{1}{M} \right)^{W-1} 
\right]  = \lambda q_{\rm o} e^{ - \frac{\lambda q_{\rm o}}{M}}.
    \label{EQ:ES}
\end{align}
From \eqref{EQ:tXm}, we have
\begin{align}
\uE[\tilde T_{(m)} ] & = \uE\left[ \frac{\tilde U_{(m)}}{B_{\rm o} \log_2 (1 + \tilde \gamma_{(m)}) } \right]  \cr
& = \frac{\uE[\tilde U_{(m)}]}{B_{\rm o}  }
\uE\left[ \frac{1}{ \log_2 (1+ \tilde \gamma_{(m)} )}
\right].
    \label{EQ:tXx}
\end{align}
Under the assumption of {\bf A2}, we have
$$
\uE[\tilde U_{(m)} \,|\, \tilde U_{(m)} \le U_{\rm max}] = \frac{1}{\mu_{\rm max}}. 
$$
In addition, since
$\ln (1+x) \ge \frac{2x}{2+x}$, $x \ge 0$, we have \eqref{EQ:L1} from
\eqref{EQ:ES} and \eqref{EQ:tXx}.
The inequality in \eqref{EQ:L1_b} is due to $\uE[\tilde U_{(m)}] = \frac{1}{\mu} \ge \uE[\tilde U_{(m)} \,|\, \tilde U_{(m)} \le U_{\rm max}] =
\frac{1}{\mu_{\rm max}}$.

\section{Proof of Lemma~\ref{L:3}}  \label{A:3}

From \eqref{EQ:EvZ}, it follows
\begin{align}
\uE [ e^{\nu  Z_{(k)}}  ] & = \frac{1}{1-z}
\uE \left[ \frac{1- z^S}{S} \right] \cr 
& = \frac{1}{1-z} \sum_{s=1}^\infty \frac{1- z^s}{s}
\frac{e^{-\bar \lambda} \bar \lambda^s}{s!} \cr 
& = \frac{e^{-\bar \lambda}}{1-z}
\left( \sum_{s=0}^\infty \frac{\bar \lambda \psi_{\bar \lambda} (s)}{(s+1)^2}  - 
\frac{\bar \lambda z \psi_{\bar \lambda z} (s)}{(s+1)^2} 
\right),
    \label{EQ:A31}
\end{align} 
where $\psi_x (s) = \frac{x^s}{s!}$.
In order to find a closed-form expression, we need to have the following 
result.

\begin{myproposition}
It can be shown that 
\begin{align} 
\frac{1}{(s+1)^2} & =  \frac{1}{(s+1) (s+2)}
+ \sum_{n=3}^\infty \frac{(n-2)!}{\prod_{i=1}^n (s+i)} \cr 
& = \sum_{n=2}^\infty \frac{(n-2)!}{\prod_{i=1}^n (s+i)} .
    \label{EQ:app_s12}
\end{align} 
\end{myproposition}
\begin{IEEEproof}
It can be shown that
\begin{align*}
\frac{1}{(s+1)^2} & = \frac{1}{(s+1) (s+2)}
+\left( \frac{1}{(s+1)^2} - \frac{1}{(s+1) (s+2)} \right) \cr
& = \frac{1}{(s+1) (s+2)}  +
\underbrace{\frac{1}{(s+1)^2 (s+2)}}_{=R_3(s)}.
\end{align*}
Then, the difference term, $R_3 (s)$, can also be written as
\begin{align*}
R_3(s)
=  \frac{1}{(s+1) (s+2) (s+3)} 
 + \underbrace{\frac{2}{(s+1)^2 (s+2) (s+3)}}_{=R_4 (s)}. 
\end{align*}
The difference term, $R_4 (s)$, can be further expressed as
\begin{align*}
R_4 (s) = 
\frac{2}{(s+1) (s+2) (s+3) (s+4)} + R_5(s),
\end{align*}
where $R_n (s)$, $n \ge 4$, can be defined as
$$
R_{n+1} (s) = R_{n} (s) - \frac{(n-2)!}{\prod_{i=1}^n (s+i)}.
$$ 
Then, after some additional manipulations, 
we can show \eqref{EQ:app_s12}.
The difference term, $R_n (s)$, becomes smaller as $n$ increases
for $s \ge 0$.
\end{IEEEproof}

In \eqref{EQ:A31}, to find each term on the right-hand side (RHS),
using \eqref{EQ:app_s12},
we can use the following expression: 
\begin{align}
\sum_{s=0}^\infty \frac{\psi_x(s)}{(s+1)^2}
& = \sum_{s=0}^\infty \frac{x^s}{(s+1)^2 s!} \cr 
& = \sum_{n=2}^\infty
\frac{(n-2)!}{x^n} \sum_{s=0}^\infty \frac{x^{s+n}}{(s+n)!}  
= \sum_{n=2}^\infty \beta_n (x). \qquad 
    \label{EQ:beta_a}
\end{align} 
Substituting \eqref{EQ:beta_a} into \eqref{EQ:A31}, 
we have \eqref{EQ:L3}.

\bibliographystyle{ieeetr}
\bibliography{mtc}

\end{document}